\documentclass[prl,showpacs,twocolumn,amsmath,amssymb]{revtex4}
\usepackage{graphicx}% Include figure files
\usepackage{dcolumn}% Align table columns on decimal point
\usepackage{longtable}

\begin{document}

\begin{abstract}

Recently (PRL 96, 106804 (2006)) it was suggested that
cobaltocene(CC) molecules encapsulated into  (7,7) carbon
nanotubes (CNT@(7,7)) could be the basis for new spintronic
devices. We show here based on impact molecular dynamics and DFT
calculations that when dynamical aspects are explicitly considered
the CC encapsulation into CNT@(7,7) does not occur, it is
prevented by a dynamic barrier mainly due to van der Waals
interactions. Our results show that CNT@(13,0) having enough axial
space for encapsulation but no enough one to allow freely rotation
of the cobaltocene molecule would be a feasible candidate to such application.

\end{abstract}


\title{\bf Cobaltocene Encapsulation Into Single-walled Carbon Nanotubes: A Molecular Dynamics Investigation}

\author{David L. Azevedo$^1$, Fernando Sato$^2$, Antonio Gomes de Sousa Filho$^3$, Douglas S. Galv\~ao$^2$}\affiliation{$^1$Departamento de F\'{\i}sica, Universidade Federal do  Maranh\~{a}o, 65080-040 S\~ao Luis, Maranh\~ao, Brazil  \\ $^2$ Instituto de F\'{\i}sica Gleb Wataghin, Universidade Estadual de Campinas, CP 6165, 13083-970 Campinas, SP, Brazil.\\ $^3$Departamento de F\'{\i}sica, Universidade Federal do Cear\'a,
60451-970 Fortaleza, Cear\'a, Brazil \\} \date{\today} \maketitle

Carbon nanotubes (CNTs) have been one of the most studied
materials in the last years. Their nanosize and diameter have been
exploited as a basis for a large variety of applications
\cite{1baughman}. They exhibit very interesting electrical and
mechanical properties. Among these properties, the ability of
encapsulating atoms and molecules \cite{2smith,3khlobystov} can
be used to engineer one or quasi-onedimensional systems.

For instance, the encapsulation of fullerene molecules into CNTs
lead to self- assembled structures varying from linear chain
(generically named peapods) \cite{2smith,3khlobystov} to very
complex ones such as multi-helices
\cite{4mickelson,5troche,6zhou}. In general, there is not a strong
electronic coupling between the fullerenes and the CNTs due to
their weak interactions, mainly van der Waals (vdW) ones
\cite{7hirahara,8Lu}. When fullerenes are replaced with structures
containing metallic atoms more complex structures can be formed
with potential significant electronic interactions between
molecules and CNTs \cite{9sceats,10ivanovskaya}. Diameter
selective effect for the encapsulation of cobaltocene molecules
into CNT has been experimentally reported \cite{11Li}.

More recently, Garc\'{\i}a-Su\'{a}rez and collaborators(GFL)
\cite{12garcia1,13garcia2,lambert2007} suggested that systems
composed of cobaltocene (CNT@(7,7) or CNT@(8,8)) could be the
basis for new spintronics devices. Cobaltocenes are molecules
composed of two aromatic pentagonal rings (C5H5) sandwiching one
cobalt atom. The authors based their conclusions on DFT (density
functional theory) calculations using the SIESTA
\cite{14soler,15siesta} code. No temperature or dynamical effects
were considered. Also, it must be considered that it is a
well-known fact \cite{16primer,17yildirim} that DFT methods do not
describe well the vdW interactions, especially in the GGA
(generalized gradient approximation) used by GFL
\cite{12garcia1,13garcia2}. In general, LDA (local density
approximation) underestimates the bong-length values while GGA
tends to overestimate them. In this sense in problems where the
vdW interactions are very important, as in the case of cobaltocene
encapsulated into CNTs, LDA would be the best choice, because it
underestimates the bond-lengths, it overestimating the covalent
aspects and, consequently, captures part of the missing vdW
interactions.

\begin{figure}[htb]
\begin{center}
\includegraphics[width=8cm]{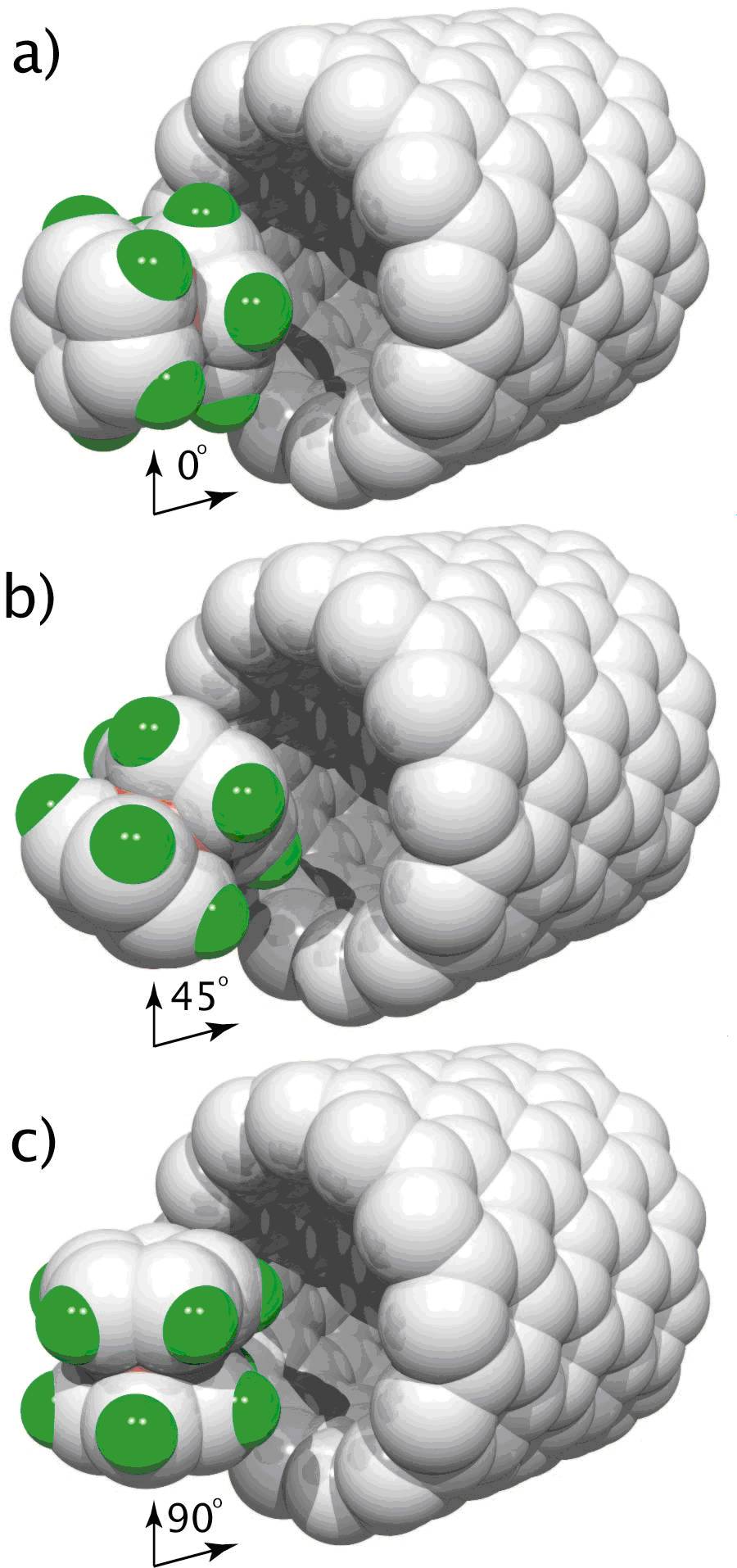}
\caption{Different initial configurations for the molecular dynamics simulations. We have considered cobaltocene configurations at parallel (a), 45$^{o}$ tilted (b), and perpendicular (c) orientations in relation to the nanotube axial axis.} \label{fig1}
\end{center}
\end{figure}

Recent classical molecular dynamics simulations
\cite{5troche,6zhou} of the encapsulation of fullerenes into CNTs
have shown that dynamical aspects and vdW interactions are of
fundamental importance. In order to determine whether this is also
the case for the encapsulation of cobaltocenes into CNTs we have
carried out extensive molecular dynamics simulations and DFT ab
initio calculations for those systems. We have investigated the
CNT@(7,7), CNT@(8,8), CNT@(13,0), and CNT@(14,0). We have
considered cases of frozen and free to relax CNTs. Our results
show that when the dynamical aspects are explicitly taken into
account, the cobaltocene encapsulation into CNT@(7,7) does not
occur, in disagreement with GFL \cite{12garcia1,13garcia2}
results. For the others investigated CNTs the encapsulation was
observed.
Initially we will discuss the results from MD
simulations. We carried out a systematic study of impulse MD
calculations, where the cobaltocene molecules, at different
relative orientations and initial velocities (1.0, 5.0, and 10
\AA/ps) are directed toward the CNTs, as schematically shown in
Figure \ref{fig1}.  The MD simulations were carried out using the
UFF (universal force field), implemented in Cerius2 package
\cite{20cerius2}. UFF contains bond stretch, bond angle bending,
inversion, torsion, rotational, and vdW terms. It has been shown
to produce accurate results for organics \cite{18rappe,19root} and
metals \cite{21rappe}. We have successfully used it to study
carbon nanotubes \cite{22legoas}, carbon nanoscrolls
\cite{23braga}, and organics over metallic surfaces
\cite{24otero}.

\begin{figure}[htb]
\begin{center}
\includegraphics[width=\columnwidth]{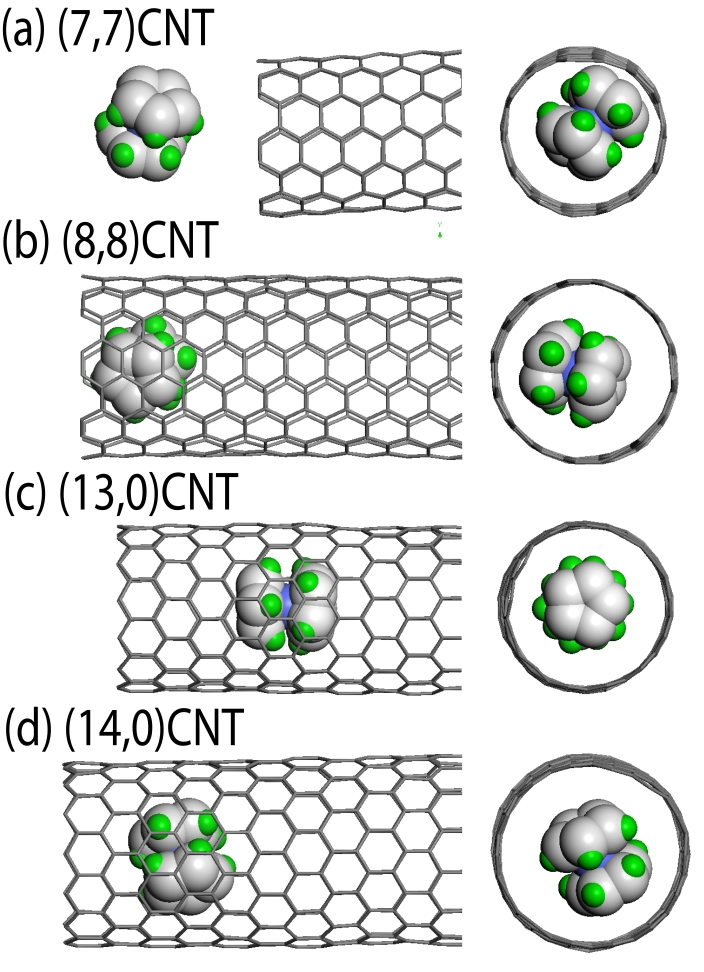}
\caption{Representative snapshots from molecular dynamics simulations for the CC encapsulation processes of a) CNT@(7,7); b) CNT@(8,8); c) CNT@(13,0); d) CNT@(14,0), respectively. For all the cases considered here (different velocities and initial cobaltocene orientations) we did not observe the cobaltocene encapsulation into CNT@(7,7) (a), while for all the other nanotubes considered it was observed (b-d).} \label{fig2}
\end{center}
\end{figure}

In Figure \ref{fig2} we present representative snapshots from MD
simulations for the different nanotubes considered here. A better
view of the process can be obtained from the movies in the
supplementary materials \cite{25epaps}.

For all the cases we investigated (different velocities and
orientations) we did not observe the cobaltocene encapsulation into
CNT@(7,7). The cobaltocene molecules, even at the most favorable
situations, are ``trapped'' at the nanotube borders (see videos 01
e 02). Although the configuration of the cobaltocene inside the
nanotube is energetically favorable, there is a dynamic barrier,
mainly due to vdW interactions, that prevents the cobaltocene
encapsulation into CNT@(7,7). We have also carried non-impulse
dynamics placing CC near CNT ends and we obtained similar results,
no encapsulation was observed for CNT@(7,7).

\begin{figure}[htb]
\begin{center}
\includegraphics[width=\columnwidth]{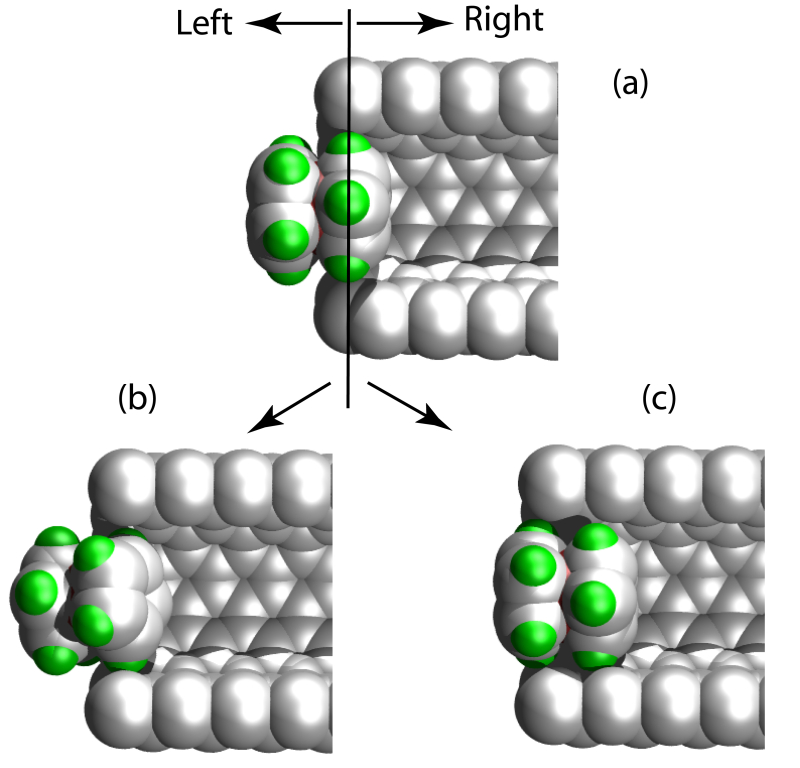}
\caption{Snapshots from rigid-body molecular mechanics simulations
of the encapsulation processes of a cobaltocene into CNT@(7,7).
Even at favorable initial configurations (a-c) the molecule never
crosses the CNT border (c).} \label{fig3}
\end{center}
\end{figure}

For the other investigated tubes (CNT@(8,8), CNT@(13,0), and CNT@(14,0)) we observed the cobaltocene encapsulation. For the cases CNT@(8,8) and CNT@(14,0), there is enough free space so the barrier for cobaltocene rotation is small. They can move almost freely. But, as well pointed out by GFL \cite{12garcia1,13garcia2} this rotation freedom will render the nanodevices useless at room temperature, since the potentially magnetic ordering would be destroyed by the available thermal energies.

On the other hand for the CNT@(13,0), there is enough axial space to allow an easy encapsulation, but not enough to allow freely rotations (see video 03). Thus, from a structural point of view our results suggest that CNT@(13,0), and not CNT@(7,7) as proposed by GFL \cite{12garcia1,13garcia2}, would be a feasible candidate for these kind of applications.

As mentioned before our conclusions are in clear disagreement with
GFL \cite{12garcia1,13garcia2} ones and the origin of these
discrepancies needs to be addressed. One obvious possibility it is
that this is just the consequence of different geometrical
results, and in this case the reliability of our molecular force
field results against the DFT-SIESTA ones needs to be established.

In order to do this we carried out a comparative study of the
geometries of the cobaltocene molecules, the nanotubes, and some
selected configurations involving cobaltocene and nanotubes. We
contrasted the results from UFF with the ones obtained using DMol3
\cite{26delley,27dmol3} and SIESTA\cite{15siesta} code. DMol3 is state of art \textit{ab initio} DFT methodology. In all our
DMol3 simulations we have used relativistic all-electron DFT total
energy approach in LDA (Wang-Perdew \cite{28pw}
exchange-correlation functional) and GGA \cite{29pbe}
approximations. A double numerical basis set with polarization
functions was also considered.

To the Siesta \cite{siesta1,siesta2} calculations we have used the standard double zeta
plus polarization(DZP) basis with an energy shift of 0.27 eV to
represent the pseudoatomic confinement for all atoms.
A cutoff of 180 Ry for the grid integration was
utilized to represent the electronic charge density for both (LDA
- CA) \cite{ca} and generalized gradient approximation (GGA - PBE)
\cite{pbe} for CC molecule and CNTs. All atoms
position were set free and  the structure was relaxed for forces
below to $0.04$ eV/\AA\ using a conjugate gradient algorithm. The
pseudopotential was constructed according to Troulier Martins
scheme.

In Tables I and II we present a summary of the main geometrical
features obtained with the different methods. As we can see there
is an excellent agreement between the geometrical data obtained
with the classical force field and the ab initio DFT ones. In
particular, although the cobaltocene's H(high) dimension
using UFF is a little off in relation to the DMol3 ones the
relevant magnitude determining the encapsulation is L, and in this
case it is in very good agreement with DMol3 results. It
should be stressed from the experimental point of view it is
difficult to determine the chirality of CNT based on the diameter
estimations. Experimental diameter estimatives from transmission
electron microscopy(TEM) measurements contain errors as large as
30\% for CNT with diameters less than 1.0 nm and about 10\% for CNT
of large diameters \cite{30qin}. Considering these error bars it is
not possible to discriminate (7,7) from (13,0) CNTs.

\begin{table}
\caption{Summary of the main geometrical features for the
investigated CNTs. The lateral (L) and vertical (H) cobaltocene
dimensions are also displayed. Results in Angstrons. }\label{tab1}
 \begin{ruledtabular}
  \begin{tabular}{ccccccc}
           & (7,7) & (8,8) & (13,0) & (14,0) & L & H \\
GGA(Dmol3) & 9.525 &  10.777 & 10.298 & 11.024 & 4.602 &  3.486 \\
LDA(Dmol3) & 9.489 &  10.753 & 10.218 & 11.023 & 4.596 &  3.360 \\
\textbf{UFF(Cerius)} & 9.470 & 10.740 & 10.160 & 10.934 & 4.535 &  3.124 \\
GGA(Siesta) & 9.631 & 10.969 & 10.329 & 11.111 & 4.442 &  3.552 \\
LDA(Siesta) & 9.540 & 10.868 & 10.226 & 11.009 & 4.452 &  3.443 \\
 \end{tabular}
\end{ruledtabular}
\end{table}

\begin{table}
\caption{Difference (in Angstroms) between the different methods for the magnitudes indicated in Table I.} \label{tab2}
\begin{ruledtabular}
\begin{tabular}{ccccccc}
  % after \\: \hline or \cline{col1-col2} \cline{col3-col4} ...
 & (7,7) & (8,8) & (13,0) & (14,0) & L & H \\
GGA/LDA(Dmol3) & 0.036 &  0.024 &  0.080 &  0.001 &  0.006 &  0.126 \\
GGA(Dmol3)/UFF & 0.055 &  0.037 &  0.138 &  0.090 &  0.067 &  0.362 \\
LDA(Dmol3)/UFF & 0.019 &  0.013 &  0.058 &  0.089 &  0.061 &  0.236 \\
GGA(Siesta)/UFF & 0.019 &  0.013 & 0.058 &  0.089 & -0.093 &  0.428 \\
LDA(Siesta)/UFF & 0.070 &  0.128 & 0.066 &  0.075 & -0.083 &  0.319 \\
GGA(Dmol3/Siesta) & -0.106 & -0.192 & -0.031 & -0.087 & 0.160 & -0.066 \\
LDA(Dmol3/Siesta) & -0.051 & -0.115 & -0.008 & 0.014 & 0.144 & -0.083 \\

\end{tabular}
\end{ruledtabular}
\end{table}

\begin{table*}
\begin{ruledtabular}
\begin{tabular}{ccccc}
            & (7,7)         & (8,8)         & (13,0)         & (14,0) \\
LDA(Dmol3)  & 4.895         & 6.159         & 5.624          & 6.429 \\
GGA(Dmol3)  & 4.923 (0.028) & 6.175 (0.016) & 5.696 (0.072)  & 6.422 (-0.007) \\
UFF(Cerius) & 4.935 (0.040) & 6.205 (0.046) & 5.625 (-0.082) & 6.399 (-0.030) \\
LDA(Siesta) & 5.088 (0.193) & 6.416 (0.257) & 5.774 (0.150)  & 6.557 (0.128) \\
GGA(Siesta) & 5.189 (0.294) & 6.527 (0.368) & 5.887 (0.263)  & 6.669 (0.240) \\
\end{tabular}
\caption{Difference (in Angstroms) between the tube diameter value
and the lateral (L) cobaltocene  dimension, for the different
methods considered here. In parenthesis are displayed the
differences relative to the LDA ones.} \label{tab3}
\end{ruledtabular}
\end{table*}

In Table III we present the differences of the ``free space" (not
considering the excluded volume due to the van der Waals
repulsions) for the different methods. Although the differences
are very small it could be argued that if we are in the limit
cases these differences could be decisive determining whether a
molecule would be encapsulated or not. In order to rule out this
possibility we run a final test where we used the SIESTA-GGA
geometries for the tube and cobaltocene molecules in a simulation
of rigid body (the geometries are keep fixed during the process of
energy minimization). Even in this limit case using the SIESTA
geometries the cobaltocene molecule is prevented to be
encapsulated by the vdW interactions (not well described in DFT
approaches) (Figure \ref{fig3}). It should stressed that
for CNT diameters considered here no significant charge transfer
was observed from SIESTA calculations \cite{9sceats}. These
results strongly indicate that the encapsulation of cobaltocene
into CNT@(7,7) is not possible, in contrast with GFL results \cite{12garcia1}.

In summary, we have investigate using impact classical molecular
dynamics and ab initio DFT methods the encapsulation processes of
cobaltocene molecules (CC) into carbon nanotubes. In contrast with
previous DFT studies \cite{12garcia1,13garcia2} our results show
that, the CC encapsulation into CNT@(7,7) it is not possible when
dynamical effects are explicity taken into account. Our results
also show that CNT@(13,0) would be a better candidate for
spintronic applications. These results also point out that
conclusions based on DFT results of geometrical configurations
where van der Waals are of central importance should be take with
caution.

\section{Acknowledgments}

The authors acknowledge financial support from Rede Nacional de Pesquisa em Nanotubos de Carbono (IMMP/MCT, IN/MCT, THEO-NANO) and the Brazilian agencies FAPESP, CNPq, CAPES and FAPEMA. Our calculations were partially performed at \textsc{LSIM} at UFMA and in \textsc{GSONM} laboratory at \textsc{IFGW-UNICAMP}. AGSF acknowledges the support from Brazilian agencies FUNCAP (grant 985/03), CNPq (grants 556549/2005-8, 475329/2006-6, 307417/2004-2).

% inseri os agradecimentos pela sugestao do Antonio Gomes

%This work was supported by the Brazilian agencies CNPq, CAPES and FAPEMA. Our calculations were partially performed at \textsc{LSIM} at UFMA and in \textsc{GSONM} laboratory at \textsc{IFGW-UNICAMP}.

\end{document}